\documentclass[useAMS,usenatbib,usegraphicx]{mn2e} %
\usepackage{amsmath,amssymb}
\usepackage{psfrag}
\usepackage{rotating}
\usepackage{pstricks}

\newcommand{\be}{\begin{equation}}
\newcommand{\ee}{\end{equation}}
\newcommand{\bea}{\begin{eqnarray}}
\newcommand{\eea}{\end{eqnarray}}
\newcommand{\beao}{\begin{eqnarray*}}
\newcommand{\eeao}{\end{eqnarray*}}
\newcommand{\dif}[2]{\frac{\partial #1}{\partial #2}}
\newcommand{\diff}[2]{\frac{\partial^2 #1}{\partial {#2}^2}}
\newcommand{\tdif}[2]{\frac{d #1}{d #2}}

\newcommand{\nn}{\nonumber}

\newcommand{\referee}{\relax}

\title[On the Eddington limit in accretion discs]{On the Eddington
limit in accretion discs}

\author[D. Heinzeller and W.J. Duschl]{D.
Heinzeller$^{1}$\thanks{E-mail:
dh@ita.uni-heidelberg.de; IMPRS Heidelberg fellow} and W.J.
Duschl$^{1,2}$\\
$^{1}$Zentrum f\"{u}r Astronomie Heidelberg, Institut f\"{u}r
Theoretische Astrophysik, Albert-Ueberle-Stra{\ss}e 2, 69120
Heidelberg, Germany;\\
\phantom{$^{1}$}{\referee now at: Institut f\"{u}r Theoretische Physik und Astrophysik, Universit\"{a}t Kiel, 24098 Kiel, Germany}\\
$^{2}$Steward Observatory, The University of Arizona, 933 North
Cherry Ave, Tucson, AZ 85721, USA}

\begin{document}

\date{Accepted $\clubsuit$. Received $\clubsuit$; in original
form $\clubsuit$}

\pagerange{\pageref{firstpage}--\pageref{lastpage}}
\pubyear{$\clubsuit$}

\maketitle

\label{firstpage}

\begin{abstract}
Although the Eddington limit has originally been derived for stars,
recently its relevance for the evolution of accretion discs has been
realized. We discuss the question whether the classical Eddington limit --
which has been applied globally for almost all calculations on accretion
discs -- is a good approximation if applied locally in the disc.
For this purpose, a critical accretion rate corresponding to this type
of modified classical Eddington limit is calculated from
{thin $\alpha$-disc models and slim disc models. We account for the
non-spherical symmetry of the disc models by computing the local upper
limits on the accretion rate from vertical and radial force equilibria
separately.}

It is shown that the results can differ considerably from the classical
(global) value: {The vertical radiation force limits the maximum
accretion rate in the inner disc region to much less than the classical
Eddington value in thin $\alpha$-discs, while it allows for significantly higher accretion
rates in slim discs.} We discuss the implications of these results
for the evolution of accretion discs and their central objects.
\end{abstract}


\section{Introduction}
Similar to the stellar case, disc accretion may be limited by
radiation pressure, counteracting gravity and viscous dissipation.
The major difference to the stellar case is the different geometry
which potentially complicates the situation considerably. While in
the stellar case, we are dealing with a -- for all practical
purposes -- spherically symmetric, i.\,e., 1-dimensional situation,
discs require an -- at least -- 2-dimensional treatment. As yet,
however, usually a quasi-stellar Eddington limit is being applied
for disc models: (1) spherical symmetry of the
system; (2) isotropic radiation; (3) homogeneous degree of
ionisation; and (4) no time dependence (stationarity). All these
approximations do not apply for the disc case, though it is not
clear {\it a priori\/} to what degree a proper treatment will alter
the resulting numbers. In addition, independent of whether we are
dealing with the stellar or the disc case, classically, the
Eddington limit uses several approximations which may or may not be
justified in all cases: (i) Thomson scattering as the sole source of
opacity; (ii) negligible gas pressure, in comparison with radiative
pressure; and (iii) no relativistic effects. It is the goal of the
present investigation to find out the importance of relinquishing
some or all of these approximations.

Over the at least two decades, a number of efforts have been carried
out to investigate the applicability of an Eddington-type limit to
the disc case: \citet*{jaros_1} and \citet*{abra_thick}, for
instance, have been working on supercritical accretion discs. As a
main result of their work, models for so-called slim and thick
accretion discs have been developed. In addition to standard
radiative cooling, in these discs also cooling by advective flows is
taken into account. This permits the critical accretion rate to
increase without rising the radiative flux accordingly. The photon
trapping mechanism \citep{begelman_1978,ohsuga_2002} amplifies this
effect even more.

\citet*{collin_2004} inspected Narrow Line Seyfert Galaxies 1 and concluded
that these accrete at super-Eddington rates, while their luminosity
stays of the order of the classical Eddington limit. The authors
proposed an additional non-viscous energy release in the gravitationally
unstable region of the disc, which emits a fraction of the optical luminosity.
Recently, \citet*{turner_2005} investigated the effect of
photon bubble instabilities by two- and three-dimensional
numerical radiation MHD calculations, finding that photon bubbles
may be important in cooling radiation-dominated accretion discs
and therefore confirming previous work by \citet*{gammie_1998},
\citet*{begelman_2002} and \citet*{ruszkowski_2003}.
In a similar approach, \citet{meyer_2005} proposed that in strong magnetic fields
under radiation pressure, discs fragment into individual magnetically aligned
columns of cool disc gas and that radiation escapes through the gaps
between them at super-Eddington luminosity.

{So far most investigations assumed a} \emph{global} limit in analogy to the classical
Eddington limit to be valid also in the context of an accretion disc.
However, \citet*{fukue_1,fukue_2} questioned this assumption for the first time {quantitatively} and
calculated a \emph{local} Eddington-type limit for
radiation pressure dominated geometrically thin discs as well as for slim discs
by applying self-similar solutions, based on the work of~\citet*{watarai}.
He found that inside a certain
\emph{critical} radius $s_\textup{cr}$, the disc remains in an
Eddington-critical state with decreasing local accretion rates
$\dot{M} = \dot{M}_\textup{input} \cdot (s/s_\textup{cr})$,
independent of the disc model. The critical radius scales with the
mass input rate, $s_\textup{cr} \approx 1.95
(\dot{M}_\textup{input}/\dot{M}_\textup{E}) \cdot s_\textup{g}$, with $\dot{M}_\textup{E}$
being the classical Eddington rate\footnote{The classical Eddington
accretion rate $\dot M_\textup{E}$ is derived by equating the disc
luminosity with the Eddington limit and deriving a mass
flow rate from it (see Sect.~\ref{sec_general_setup})} and $s_\textup{g}$ the gravitational
radius of the central black hole.

In the present paper, we calculate the analog to the Eddington limit
(resp., the critical accretion rate) for a classical thin disc model
of the \citet*{shakura_sunyaev} variety and for a slim disc model of
the \citet*{abra_slim} type. In both descriptions, a standard
$\alpha$-viscosity is applied, therefore restricting the models to the
non-selfgravitating case. We relinquish the above approximations
(1), (2), (3), (i), and (ii), but keep the approximation of
non-relativistic stationarity.

In Sect.~\ref{sec_model_setup}, we provide the models and
derive the equations for the critical accretion rates {for
both disc types and directions (vertical, radial) separately}. In
Sect.~\ref{sec_results}, the results are presented and compared to
the classical Eddington accretion rate. {Based upon that, a
locally Eddington limited disc model is proposed.}
Section~\ref{sec_discussion} is dedicated to the discussion of our results, and
Sect.~\ref{sec_conclusion} to an outlook on the possible effects on
the evolution of accretion discs.

\section{Models}\label{sec_model_setup}
The original physical reasoning for the Eddington limit is an
equilibrium of the gravitational force $F_\textup{g}$ and the radiative force
$F_\textup{r}$. Contrary to a spherically symmetric body like a star, an
accretion disc has to be treated as at least a two-dimensional
object. Therefore, one can define two different ``Eddington limits'',
corresponding to the equilibria in the vertical ($z$) and radial
(equatorial, $s$) direction:
\beao
F^{(z)}_\textup{tot}(s)&=&F^{(z)}_\textup{r}(s) - F^{(z)}_\textup{g}(s) \ = \ 0\\[3mm]
F^{(s)}_\textup{tot}(s)&=&F^{(s)}_\textup{r}(s) - F^{(s)}_\textup{g}(s) + \ldots \ = \ 0
\eeao
For a rotating viscous disc, further contributions to the total
force have to be included. Moreover, the disc Eddington limit is no
longer independent of the distance from the center of the object, as
is true for the stellar case, leading to \emph{local},
position-dependent Eddington limits.

This section is divided into three parts: The first
part introduces the general set-up which applies to thin
and slim discs likewise, while the second and third part
focus on thin and slim disc models, respectively.

\subsection{General set-up}\label{sec_general_setup}
In this contribution, we use a cylindrical coordinate system
$\{ s, \varphi, z \}$ with the distance $r$ to the origin, $r^2 = s^2 + z^2$.

The critical accretion rate at a given position $s$ is denoted by
$\dot{M}_\textup{crit} = \dot{M}_\textup{crit}(s)$, while the classical Eddington accretion rate
is identified as $\dot{M}_\textup{E}$.

If one naively identifies the classical Eddington luminosity
\be
L_\textup{E} = 4 \pi c\,\frac{G M}{\kappa_\textup{es}}
\ee
($\kappa_\textup{es} = 0.4\,\textup{cm}^2\textup{g}^{-1}$ for
Thomson-scattering) with the total energy output of an accretion disc
of
\be
L_\textup{E} = \eta \cdot \frac{G M \dot{M}_\textup{E}}{s_\textup{i}}\,,
\ee
one may derive a quasi-classical Eddington accretion rate
\be
\dot{M}_\textup{E} = \frac{4}{\eta} \cdot \pi c \frac{s_\textup{i}}{\kappa_\textup{es}}\,.
\ee
Here, $s_\textup{i}$ denotes the disc's inner radius;
the factor $\eta$ takes care of the coupling between the
innermost disc regions and the accreting object. In
this paper, we use $\eta = 1/2$. As long as $\eta$ is of order
unity, our conclusions do not depend on its exact value.

We assume an optically thick situation for all our calculations.
This, together with the assumption that selfgravity plays no role,
has to be justified afterwards.

In contrast to the classical approach, we allow for opacity sources
other than Thomson scattering and use an analytic interpolation
formula (Gail, priv.\,comm.{\referee; for a similar approach, see also \citet*{bell_1994}}):
\begin{eqnarray}
\frac{1}{\kappa_\textup{in}}&=&
\left[\frac{1}{\kappa_\textup{ice}^4} +
\frac{\left(3\,000\,\mathrm{K}\right)^{10}}{\left(3\,000\,\mathrm{K}\right)^{10}
+ T^{10}}\cdot
\frac{1}{\kappa_\textup{ice, evap}^4 + \kappa_\textup{dust}^4}\right]^{1/4}\nonumber\\
&& + \left[\frac{1}{\kappa_\textup{dust, evap}^4 +
\kappa_\textup{mol}^4 + \kappa_\textup{H$^{-}$}^4}
+\frac{1}{\kappa_\textup{atom}^4 + \kappa_\textup{e$^{-}$}^4}\right]^{1/4}
\label{eqn_kappa_interpol}
\end{eqnarray}
{\referee The individual contributors $\kappa_i$ are approximated by
\begin{equation}
\kappa_i = \kappa_{0,i} \cdot T^{\kappa_{T,i}} \cdot \rho^{\kappa_{\rho,i}}
\end{equation}}%
and are compiled in Table~\ref{tab_kappa_interpol}.\footnote{\referee Slight
differences, for instance for $\kappa_\textup{ice, evap}$ by a factor of $2$,
between \citet[Table~3]{bell_1994} and our values, are due to somewhat
different ways of combining the individual contributors. For the purpose
of the present investigation, these differences are of minor importance}
{\referee However,} $\kappa_\textup{in}$ is only meant as a handy interpolation. In particular in the
transition regions between individual contributions, one has to be aware of this
and should refrain from a physical over-interpretation there.
\begin{table*}
\caption{Interpolation of the opacity: set of parameters (in cgs-units)}\label{tab_kappa_interpol}
{\small\begin{tabular}{lcccc}
\hline
\rule[1mm]{0mm}{2mm}\textbf{Contributor $\mathbf{i}$} & \textbf{Symbol} &
$\mathbf{\kappa_{\textup{i},0}}$
    & $\mathbf{\kappa_{\textup{i},\rho}}$ & $\mathbf{\kappa_{\textup{i},T}}$ \\ \hline
\rule[1mm]{0mm}{2mm}Dust with ice mantles & $\kappa_\textup{ice}$ & $2.0 \cdot 10^{-4}$ & $0$ & $2$\\
Evaporation of ice & $\kappa_\textup{ice, evap}$ & $1.0 \cdot 10^{16}$ & $0$ & $-7$\\
Dust particles & $\kappa_\textup{dust}$ & $1.0 \cdot 10^{-1}$ & $0$ & $1/2$\\
Evaporation of dust particles & $\kappa_\textup{dust, evap}$ & $2.0 \cdot 10^{81}$ & $1$ & $-24$\\
Molecules & $\kappa_\textup{mol}$ & $1.0 \cdot 10^{-8}$ & $2/3$ & $3$\\
Negative hydrogen ion & $\kappa_\textup{H$^{-}$}$ & $1.0 \cdot 10^{-36}$ & $1/3$ & $10$\\
Bound-free and free-free-transitions & $\kappa_\textup{atom}$ & $1.5 \cdot 10^{20}$ & $1$ & {\referee $-5/2$}\\
Electron scattering & $\kappa_\textup{e$^{-}$}$ & $0.348$ & $0$ & $0$\\ \hline
\end{tabular}}
\end{table*}

We also take into account gas pressure, giving the total pressure as
\be
p_\textup{tot} = p_\textup{gas} + p_\textup{rad} = \frac{2 \rho
k_\textup{B} T}{m_\textup{H}} + \frac{4 \sigma T^4}{3c}\,.
\ee
A parameter $\beta$ describes the contribution of gas to the total
pressure in a standard way:
\be
\beta = \frac{p_\textup{gas}}{p_\textup{tot}}\ \Longrightarrow\
\frac{1-\beta}{\beta} = \frac{2 \sigma T^3 m_\textup{H}}{3c\rho
k_\textup{B}}\label{eqn_beta_def}
\ee
A main point of interest is the total luminosity of an Eddington
limited accretion disc. We calculate the disc's luminosity by
\be
L_\textup{disc} = 2 \cdot 2\pi \cdot \int_{s_\textup{i}}^{s_\textup{o}}ds\,\sigma
T_\textup{eff}^4\,,
\label{eqn_def_total_lum}
\ee
where $T_\textup{eff}$ stands for the effective temperature (surface
temperature) of the accretion disc at radius $s$, which is related
to the central temperature $T_\textup{c}$ by the {optical depth} $\tau = \kappa \rho h$,
\be
\sigma T_\textup{eff}^4 = \frac{16}{3\tau} \sigma
T_\textup{c}^4\,,
\label{eqn_def_t_eff}
\ee
replacing the vertical stratification by a one-zone approximation.

Since our models take advantage of the $\alpha$-description for the
viscosity, the mass of the disc has to be small compared to the
mass $M$ of the central object \citep*{duschl_strittmatter_biermann}.
For later comparison with the central mass, we introduce the disc's mass
\be
M_\textup{disc} = \int_{s_\textup{i}}^{s_\textup{o}} \Sigma 2 \pi s\,ds
\label{eqn_def_M_disc}
\ee
with the surface density $\Sigma = 2 \rho h$. The height $h$
of the disc is defined by the vertical hydrostatic equilibrium, $p/h = \rho\,g_{\referee z}$,
where the mass density $\rho$ and the pressure $p$ are measured in
the disc plane ($z=0$) and the gravitational acceleration $g_z$ is taken from the disc surface ($z=h$).
\subsection{Geometrically thin accretion discs}
Geometrically thin models have been the most popular
description of accretion discs in the last 30 years. Based upon the
assumption of a negligible height $h$ of the disc relative to the
radius $r$ and neglecting advective cooling,
this model is simple but powerful for many applications. We derive
the critical accretion rate separately for the vertical and radial
direction by assuming a balance of force between the out-dragging
radiative force $F_\textup{r}$ and the attracting forces, like the
gravitational force $F_\textup{g}$.
\subsubsection{Vertical direction}
In this case, we consider a particle with mass $m_\textup{H}$ at the surface $h$
of the disc. This particle is attracted by the vertical
component of the gravitational force
\be
F_\textup{g}^{(z)} = \frac{G M m_\textup{H}}{s^2} \cdot \frac{h}{s}\,,
\ee
while it is accelerated outwards by the radiative force
\be
F_\textup{r}^{(z)} = \frac{\kappa m_\textup{H}}{c} \sigma T_\textup{eff}^4\,,
\label{eqn_fr_thin_disc}
\ee
according to the Stefan-Boltzmann law. Using
\be
\sigma T_\textup{eff}^4 = \frac{3}{8\pi} \frac{G M \dot{M}}{s^3} f
\ee
and equating $F_\textup{g}^{(z)}$ and $F_\textup{r}^{(z)}$,
we obtain an equation for the \emph{local} critical accretion rate:
\be
\dot{M}_\textup{crit}(s)= \frac{8 \pi}{3} \frac{c}{\kappa} \cdot
\frac{h}{f}
\label{eqn_thin_vert_mp}
\ee
Here, $f$ is a function describing the radially inner boundary
condition. A torque-free condition, for instance, leads to $f = 1 -
\sqrt{s_\textup{i}/s}$.

Equation~\eqref{eqn_thin_vert_mp} changes the quality of the radial mass flow
rate. While in standard disc theory, it is a parameter, it now
becomes a local solution of our criticality assumption.

By means of the remaining disc equations and \eqref{eqn_beta_def},
we solve this set of equations for $\beta$ and finally end up with
$\beta = 3/4$, independent of a specific choice of the opacity
$\kappa$. This leads to
\be
\dot{M}_\textup{crit}(s) = \frac{1}{f} \cdot \frac{8 \pi}{3} \cdot
\frac{8^{3/8}}{3^{1/2}} \cdot \frac{k_\textup{B}^{1/2} c^{5/4}}{m_\textup{H}^{1/2}
\alpha^{1/8} \kappa^{9/8} \sigma^{1/8} \Omega_\textup{K}^{7/8}}\,,
\label{eqn_result_mpcrit_thin}
\ee
where $\Omega_\textup{K}$ stands for the Keplerian angular velocity. Contrary
to the solution for $\beta$, the critical accretion rate depends on
the opacity.
\subsubsection{Radial direction}\label{sec_thin_radial}
The calculation of the critical accretion rate given by a balance of
radial forces is less straight-forward then for the vertical
direction. We consider a particle in the equatorial plane of the
disc, at $z=0$. So, the attracting gravitational force is given by
\be
F_\textup{g}^{(s)} = \frac{G M m_\textup{H}}{s^2}\,.\label{eqn_Fg_radial_thin}
\ee
In contrast to our previous derivation, the particle is located
inside the disc and no longer at its surface. Assuming our particle
being located at a specific position $s$ in an optically thick
medium, additionally to the out-dragging radiation originating from
the inner shell at $s-ds$ we have to take into account also the
radiation of the outer shell at $s+ds$ pointing inwards. In brief,
we have to apply the radial difference of the radiative force instead
of its absolute value. Furthermore, we have to consider the
centrifugal force due to azimuthal motion and, coupled with it, the
viscous forces. Putting all together, this leads us to the radial
component of the Navier-Stokes equation, which is given by
{\bea
F_\textup{tot}^{(s)}&=& m_H \Omega_K^2 s - \frac{m_\textup{H}}{\rho} \cdot
\dif{p_\textup{tot}}{s} -\frac{m_\textup{H}}{2} \dif{v_{s}^2}{s}\nn\\
&&+ m_H \nu \diff{v_{s}}{s} - m_H \nu \frac{v_{s}}{s^2} - \frac{G M m_\textup{H}}{s^2} = 0 \,.\label{eqn_Ftot_radial_thin}
\eea}
By means of the disc equations, the condition of a vanishing total force
yields a differential equation for $\beta$.
{In the Keplerian case, the centrifugal and gravitational force (i.\,e., the first
and the last term of~\eqref{eqn_Ftot_radial_thin}) cancel each other,
leaving the remainder of \eqref{eqn_Ftot_radial_thin} to be solved.}

Independent of the chosen boundary values, its solution
leads to supersonic radial motion, $|v_{s}| > c_\textup{s} {= (p_\textup{tot}/\rho)^{1/2}}$, which
contradicts the thin disc model.
Hence, we investigate the corresponding value of the
accretion rate under the condition $v_{s} = A c_\textup{s}$, $|A| < 1$,
which can subsequently be expressed as
\be
\dot{M}_\textup{crit}(s)= \frac{32 \pi c}{3 \kappa} \cdot \frac{A}{\alpha} \cdot \frac{s}{(1-\beta)}\,.
\label{eqn_thin_radial_vrad_lim}
\ee
\subsection{Slim accretion discs}
{\citet{abra_slim} extended the model of supercritical accretion discs by \citet{jaros_1} to
slim accretion discs with $h \lessapprox s$. Since the height $h$ is no
longer negligible compared to the equatorial distance $s$, the
radial momentum equation has to be solved in a more precise way,
leading to a non-Keplerian rotation of the disc. Furthermore, beside
of the standard cooling by radiation, also cooling by advective mass
flows becomes important, requiring a more detailed treatment of the
energy (transport) equation.

As the vertical disc structure is known only
approximately, the authors used three correcting factors
($B_1$, $B_2$ and $B_3$), which are all of the order unity, to
correct the results of the vertical integration:\footnote{These
parameters where introduced originally by~\citet{paczynski_2}. However,
the notation of the parameters differs slightly between these
two publications. In this paper, we follow \citet{abra_slim}}
{\begin{itemize}
\item $B_1$ takes into account that the height of the
disc is no longer negligible compared to its radius,
$(s^2+h^2)^{3/2} \approx B_1 \cdot s^3$. In the model,
we use a Newtonian potential; combined with the assumption of
hydrostatic equilibrium, this leads to
\be
\frac{p_\textup{tot}}{\rho} = \frac{GM \cdot h^2}{(s^2+h^2)^{3/2}}
\approx \frac{GM \cdot h^2}{B_1 \cdot s^3}
= \frac{\Omega_\textup{K}^2 h^2}{B_1}\,.
\ee
\item $B_2$ is a measure of how efficiently the viscously dissipated
energy is converted into radiation (rather than radially advected),
\be
F_\textup{r}^{(z)} = B_2 \frac{m_\textup{H}}{\rho}{p_\textup{rad}}{h}\,.
\ee
\item $B_3$ measures the efficiency of the advective energy transport
in the energy transport equation and as such is part of modeling
the advective processes.
\end{itemize}}}
\subsubsection{Vertical direction}
The resulting set of disc equations is given in
\citet[p.~648]{abra_slim}. Starting from there and using the
parameter $\beta$ from~\eqref{eqn_beta_def}, we derive two equations
for the three unknowns $\beta$, $h$ and $l$, where $l$ is the local
specific angular momentum. The missing third
equation is again given by the Eddington condition of a vanishing
total force:
\be
F_\textup{tot}^{(z)} = F_\textup{r}^{(z)} + F_\textup{g}^{(z)} = 0
\ee
With
\be
F_\textup{g}^{(z)} = \frac{G M m_\textup{H}}{(s^2+h^2)} \cdot
\frac{h}{\sqrt{s^2+h^2}} \approx \frac{G M m_\textup{H} h}{B_1 s^3}
\ee
in terms of the slim disc description, and
\bea
F_\textup{r}^{(z)}&=&B_2 \frac{m_\textup{H}}{\rho}{p_\textup{rad}}{h} \ =\ (1 - \beta)
B_2 \frac{m_\textup{H} p_\textup{tot}}{h \rho}\nonumber\\
&=&(1-\beta) \frac{B_2}{B_1} \frac{G M m_\textup{H} h}{s^3}\,,
\eea
the Eddington condition becomes
\be
\beta = 1 - \frac{1}{B_2}\,.
\label{eqn_beta_slim_vert}
\ee
Also in this model, the ratio of gas pressure to total pressure
remains constant and is determined by the choice of the parameter
$B_2$. The critical accretion rate as well as the other disc quantities
can be calculated by solving the two (coupled) differential equations
numerically (Appendix~\ref{app_slim_vert}).

\subsubsection{Radial direction}
We find that similar to the thin disc case presented in
Sect.~\ref{sec_thin_radial}, the radial Eddington limit in slim
accretion discs leads to supersonic radial velocities, $|v_{s}| >
c_\textup{s}$, which cannot be the case for (stationary) slim discs.
Therefore, we follow the same idea as in Sect.~\ref{sec_thin_radial}
and require $v_{s} = A c_\textup{s}$ with $|A| < 1$.
In terms of the slim disc model, we obtain
\be
v_{s} = \frac{s \alpha \Omega_\textup{K}^2 h^2}{B_1 B_2 (l-l_0)}\,.
\label{eqn_def_vs}
\ee
The condition $v_{s} = A c_\textup{s}$ leads to
\be
h = B_2 \sqrt{B_1} \cdot \frac{A}{\alpha} \cdot
\left(\frac{l}{l_\textup{K}} - \frac{l_0}{l_\textup{K}}\right) s\,.
\label{eqn_slim_rad_GL_1}
\ee
Here, $l_\textup{K}$ and $l_0$ are the local Keplerian specific angular
momentum and the specific angular momentum of the disc's innermost orbit, respectively.

Starting from the radial momentum equation and the energy equation
for slim discs, we can derive two additional differential equations
for the unknowns $h$, $\beta$ and $l$. Together with
~\eqref{eqn_slim_rad_GL_1}, the system is closed and can be solved
numerically. The set of equations is shown in
appendix~\ref{app_slim_rad}.
\section{Results}\label{sec_results}
As can be seen from the above equations, the critical accretion
rates depend on the distance $s$ from the central object. We wish to
emphasize that these accretion rates represent the \emph{local}
maximum value of mass which can be transferred inwards at a specific
distance from the center in a stationary accretion state.

For the following discussion of our results, we define a
\emph{reference model}: The central object is assumed to be a
stellar black hole with $M = 10 M_{\odot}$, resulting in an inner
radius of $s_\textup{i} = 6GM/c^2 = 8.86\cdot 10^{6}\,\textup{cm}$. The outer
radius $s_\textup{o}$ is set to $5 \cdot 10^4 s_\textup{i}$. In this environment, the
classical Eddington rate is $\dot{M}_\textup{E} =1,67\cdot
10^{19}\,\textup{g}\,\textup{s}^{-1} = 2.65 \cdot 10^{-7}
M_\odot\,\textup{a}^{-1}$. The viscosity parameter $\alpha$ is set
to $0.1$. {A torque-free boundary condition is chosen at the
inner disc radius $s=s_\textup{i}$: $f = 1 - \sqrt{s_\textup{i}/s}$.}

{In the following figures, the curves are labeled with
the corresponding disc model (\texttt{thin}, \texttt{slim}),
local Eddington limit (\texttt{vertical}, \texttt{radial}) and
opacity (\texttt{el. scat.}: $\kappa = \kappa_\textup{es}$,
\texttt{int. opac.}: $\kappa = \kappa_\textup{in}$).}
\subsection{Thin accretion discs}\label{sec_results_thin_discs}
\begin{figure}
\psfrag{s/si}{$\textstyle s/s_\textup{i}$}
\psfrag{h/s}[c][c]{\begin{rotate}{-90}$\textstyle h/s$\end{rotate}}
\includegraphics[clip,width=\columnwidth]{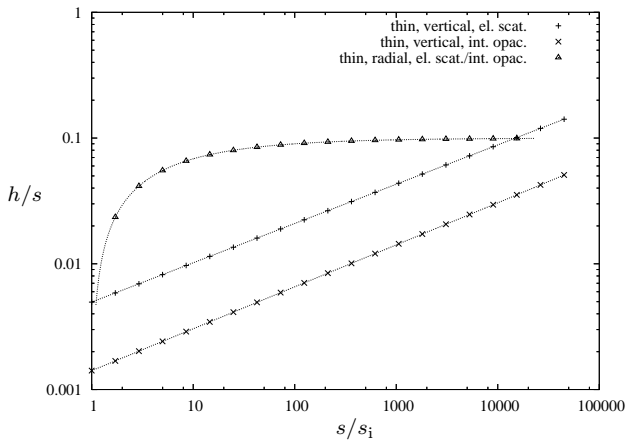}
\caption{Relative height $h/s$ for the vertical Eddington limit and
for the radial limit in standard $\alpha$-discs}\label{fig_thin_h}
\end{figure}

\begin{figure}
\psfrag{s/si}{$\textstyle s/s_\textup{i}$}
\psfrag{beta}[c][c]{\begin{rotate}{-90}$\textstyle \beta$\end{rotate}}
\includegraphics[clip,width=\columnwidth]{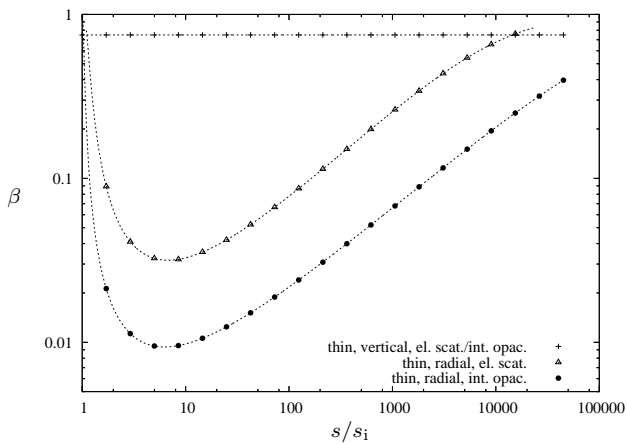}
\caption{$\beta = p_\textup{gas}/p_\textup{tot}$ for the vertical
Eddington limit and for the radial limit in standard
$\alpha$-discs}\label{fig_thin_eta}
\end{figure}

\begin{figure}
\psfrag{s/si}{$\textstyle s/s_\textup{i}$}
\psfrag{dM/dM_E}[c][c]{\begin{rotate}{-90}$\textstyle
\dot{M}_\textup{crit}/\dot{M}_\textup{E}$\end{rotate}}
\includegraphics[clip,width=\columnwidth]{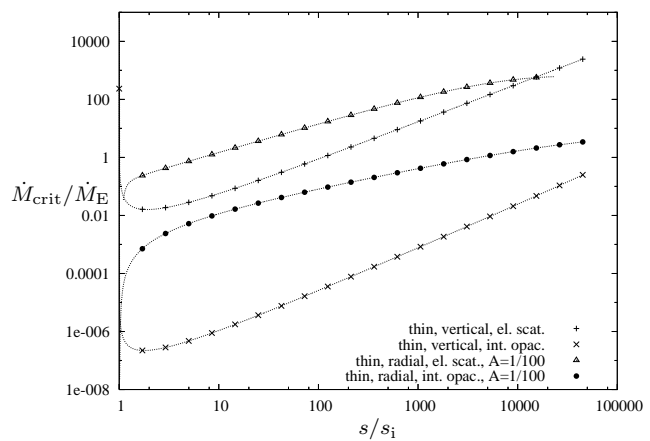}
\caption{Critical accretion rate $\dot{M}_\textup{crit}$ in units of the classical
Eddington rate $\dot{M}_\textup{E}$ for the vertical Eddington limit and for
the radial limit in standard $\alpha$-discs}\label{fig_thin_Mp}
\end{figure}

Figures~\ref{fig_thin_h}--\ref{fig_thin_Mp} present the results for the
vertical and the radial direction -- in the latter case, the parameter
$A$ is set to $0.01$, corresponding to $h/s \approx 0.1$ as an upper
limit for thin discs. {This choice is somewhat arbitrary, but
provides a resonable upper limit for the radial velocities in thin discs.
According to \eqref{eqn_thin_radial_vrad_lim}, the allowed mass accretion
rate scales linearly with the parameter $A$.}

Figure \ref{fig_thin_h} shows that the height of Eddington limited
discs decreases steadily with the radius $s$, while it remains
constant for a large range in the radial limit. This is a direct
consequence of the gas-to-total-pressure ratio $\beta$, which is
constant only for the vertical Eddington limit
(Fig.~\ref{fig_thin_eta}). In the radial limit, $\beta$ decreases
with $s$ by two orders of magnitude, causing the radiation pressure
to exceed the gas pressure and to expand the disc vertically. This
leads to higher surface densities and radial velocities, and
therefore to higher critical accretion rates, as it can be seen from
Fig.~\ref{fig_thin_Mp}. Furthermore, we observe a notable influence
of the opacity on the results: For $\kappa_\textup{es}$, the
critical accretion rates are much higher than for
$\kappa_\textup{in}$, reaching a value of approximately the
classical Eddington rate $\dot{M}_\textup{E}$ in the inner part of the disc.
{According to \eqref{eqn_fr_thin_disc} and the fact that in thin
discs all energy released by viscous processes is transported by radiation,
the radiation force depends linearly on $\kappa$, and $\dot{M}_\textup{crit} \propto \kappa^{-1}$,
see~\eqref{eqn_thin_vert_mp}.}

{For some parameter ranges we find rather large differences between
the Eddington limits derived following the standard approach of taking into
account electron scattering as the only absorbing process ($\kappa_\textup{es}$),
and those which allow for additional absorption processes ($\kappa_\mathrm{in}$).
{\referee In the present calculations, bound-free and free-free processes
(contributor $\kappa_\textup{atom}$ in~\eqref{eqn_kappa_interpol})
dominate the opacity $\kappa_\mathrm{in}$ and alter it significantly.}
{\referee The resulting differences in the Eddington limits} are of relevance
and emphasize the importance of taking into
account absorption processes other than Thomson scattering in the disc case,
i.\,e., of using consistent opacities.

The changes in the inner few $s_\textup{i}$ is due to the boundary condition
$f$ at the inner radius: For $s \to s_\textup{i}$ the boundary function $f$
tends to zero. As $\dot{M}_\textup{crit}$ scales with $f^{-1}$
(see~\eqref{eqn_result_mpcrit_thin}), the critical accretion rate
formally diverges. This, however, has to be taken as a technical artefact:
In reality, the stress at the inner boundary will always be nonzero, leading
to finite values for $\dot{M}_\textup{crit}$. However, $f$ may be still be small and
therefore theoretically allowing for higher accretion rates than at the minimum at $s_\textup{min} \approx 2 s_\textup{i}$.
As long as no vertical mass inflows to the disc are present,
the \emph{real} accretion rate $\dot{M}$ will stay at its minimum value
$\dot{M}_\textup{crit}(s_\textup{min})$ and the disc will be in a sub-Eddington state
in terms of our local Eddington-limit for $s_\textup{i} < s < s_\textup{min}$.

For the vertical Eddington limit and also for the radial limit,
the condition of a non-self\-gravitating disc holds,
{$M_\textup{disc} \ll M$}. Also the assumption
of optical thickness is fulfilled for both $\kappa_\textup{es}$ and
$\kappa_\textup{in}$,
{\[
\int_0^{z=h} \rho \kappa\,dz \approx \frac{1}{2} \Sigma \kappa > 1\,.
\] }
\subsection{Slim accretion discs}\label{sec_results_slim_discs}
For the model calculations, we choose the three parameters $B_1$,
$B_2$, and $B_3$ as follows:
{\begin{itemize}
\item The slim disc model is valid for $h \lessapprox s$,
i.\,e. $(s^2 + h^2)^{3/2} = B_1 s^3 \lessapprox 2^{3/2} s^3$, making $B_1 = 2$ a reasonable choice.
\item For the vertical Eddington limit, $\beta = 1 - 1/B_2$. In order to
compare our results directly to thin accretion discs with $\beta = 3/4$, we set $B_2=4$.
\item We follow~\citet{paczynski_2} and set $B_3 = 1/2$.
\end{itemize}}
In order to solve the sets of differential equations, we have to
define boundary conditions. Due to numerical reasons,
this has to be done at the outer boundary. We set $l(s_\textup{o}) =
l_\textup{K}(s_\textup{o})$ and $\beta(s_\textup{o})=3/4$ for the radial limit. With $A = 1/30$
(for slim discs, radial velocities are usually higher than
for thin discs) and~\eqref{eqn_slim_rad_GL_1}, this leads to $h(s_\textup{o}) =
2 s_\textup{o}$. {Also here, the choice of $A$ is somewhat arbitrary. Since the dependency
of the allowed mass accretion rates on $A$ cannot be seen directly from the equations,
we also compute the case $A=1/20$.} For
the vertical limit, we set $l(s_\textup{o}) = l_\textup{K}(s_\textup{o})$ and $h(s_\textup{o}) = 2 s_\textup{o}$.
For comparison, we also compute the vertical limit with $h(s_\textup{o}) =
s_\textup{o}$. The results are shown in Figs.~\ref{fig_slim_l}--\ref{fig_slim_Mp}.
\begin{figure}
\psfrag{s/si}{$\textstyle s/s_\textup{i}$}
\psfrag{l/l_K}[c][c]{\begin{rotate}{-90}$\textstyle l/l_\textup{K}$\end{rotate}}
\includegraphics[clip,width=\columnwidth]{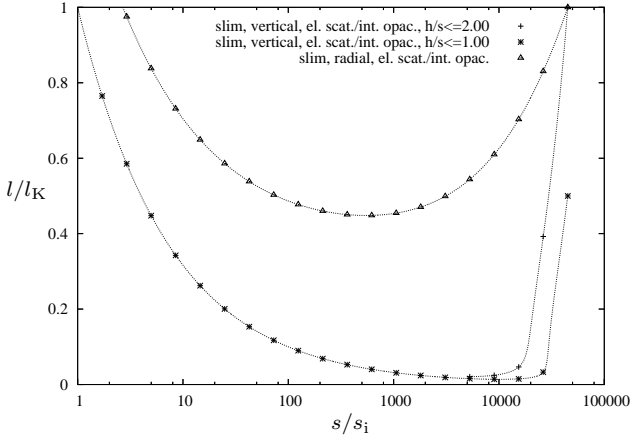}
\caption{Angular momentum $l$ in units of the Keplerian angular
momentum $l_\textup{K}$ for the vertical Eddington limit and for the radial
limit in slim discs}\label{fig_slim_l}
\end{figure}
\begin{figure}
\psfrag{s/si}{$\textstyle s/s_\textup{i}$}
\psfrag{h/s}[c][c]{\begin{rotate}{-90}$\textstyle h/s$\end{rotate}}
\includegraphics[clip,width=\columnwidth]{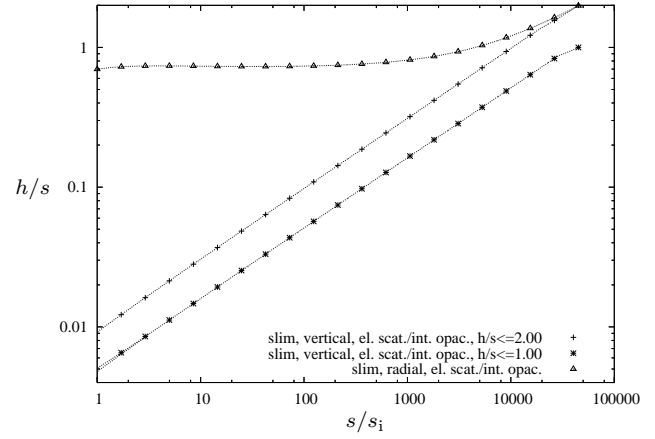}
\caption{Relative height $h/s$ for the vertical Eddington limit and
for the radial limit in slim discs}\label{fig_slim_h}
\end{figure}
\begin{figure}
\psfrag{s/si}{$\textstyle s/s_\textup{i}$}
\psfrag{beta}[c][c]{\begin{rotate}{-90}$\textstyle \beta$\end{rotate}}
\includegraphics[clip,width=\columnwidth]{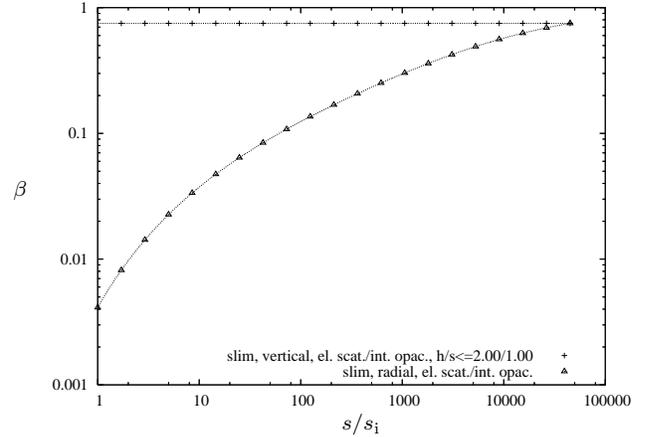}
\caption{$\beta = p_\textup{gas}/p_\textup{tot}$ for the vertical
Eddington limit and for the radial limit in slim
discs}\label{fig_slim_eta}
\end{figure}
\begin{figure}
\psfrag{s/si}{$\textstyle s/s_\textup{i}$}
\psfrag{dM/dM_E}[c][c]{\begin{rotate}{-90}\raisebox{2mm}{$\textstyle
\dot{M}_\textup{crit}/\dot{M}_\textup{E}$}\end{rotate}}
\includegraphics[clip,width=\columnwidth]{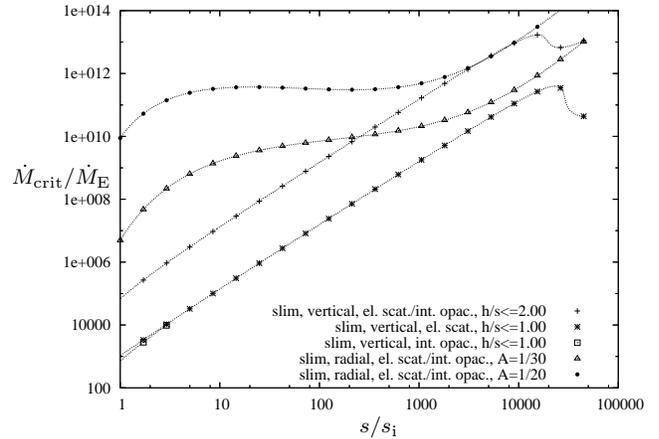}
\caption{Critical accretion rate $\dot{M}_\textup{crit}$ in units of the classical
Eddington rate $\dot{M}_\textup{E}$ for the vertical Eddington limit and for
the radial limit in slim discs}\label{fig_slim_Mp}
\end{figure}

As it can be seen from Fig.~\ref{fig_slim_l}, the angular momentum
drops with respect to the local Keplerian value $l_\textup{K}(s)$ for both
the radial and the vertical limit. For the latter one, the decrease
is extremely steep, resulting in $l(s) = (1+\varepsilon) l_0$,
$\varepsilon \ll 1$. For $s \to s_\textup{i}$, the angular momentum reaches
its Keplerian value in the vertical limit, while it overshoots this
limit at $s\approx s_\textup{i}$ in the radial limit.

Figure~\ref{fig_slim_h} shows the behaviour of the disc's relative
height {$h/s$ for local equilibria of forces in the vertical
and radial direction, respectively.} While this ratio scales with
{$s$} in the vertical {limit}, it adopts a constant value about unity in the radial
{limit}. In analogy to thin discs, this corresponds to a decrease of
$\beta$ in the latter case for smaller radii
(Fig.~\ref{fig_slim_eta}) and to higher accretion rates with respect
to the vertical limit (Fig.~\ref{fig_slim_Mp}). The critical
accretion rates exceed the limits for thin discs
(cf. Sect.~\ref{sec_results_thin_discs}) significantly.

We want to point out that the results are almost independent of the
opacity, although the absolute values of $\kappa$ differ considerably
(Fig.~\ref{fig_kappa_vgl}). This becomes plausible when regarding
the energy equation \citep[(12)]{abra_slim}: For high accretion
rates, the radiative processes are negligible compared to the
advective flows.
\begin{figure}
\psfrag{s/si}{$\textstyle s/s_\textup{i}$}
\psfrag{kappa [cm^2/g]}[r]{\begin{rotate}{-90}\raisebox{2mm}{$\textstyle
\kappa\,[\textup{cm}^2/\textup{g}]$}\end{rotate}}
\includegraphics[clip,width=\columnwidth]{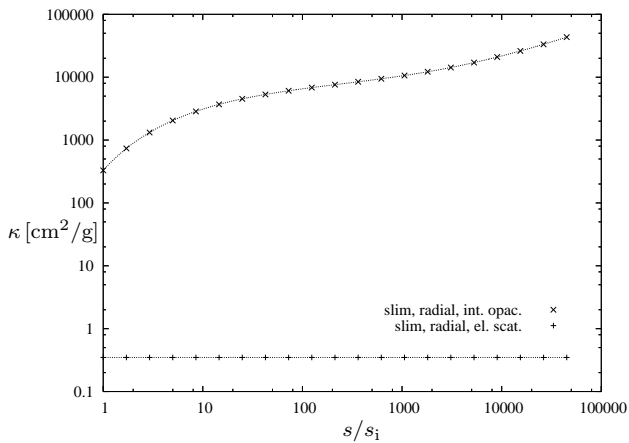}
\caption{$\kappa_\textup{es}$ and $\kappa_\textup{in}$ for slim discs in
the radial limit}\label{fig_kappa_vgl}
\end{figure}

In all cases, the assumption of large optical thickness is
fulfilled. By computing realistic disc models
(see Sect.~\ref{sec_edd_lim_disc} for a more detailed
discussion), we find that also
the condition of a negligible disc mass is fulfilled
as long as the mass input rate at the outer boundary
$\dot{M}(s_\textup{o})$ stays at a reasonable value below $10^5 \dot{M}_\textup{E}$.
\subsection{An Eddington limited thin disc}\label{sec_edd_lim_disc}
We have seen in Sect.~\ref{sec_results_thin_discs} that the
Eddington limit poses strong limitations on the mass inflow rate in
classical thin accretion discs. Contrary to the above sections where
we displayed the disc properties for critical accretion rates at
every radius, we now construct a stationary thin disc model by taking
into account the local Eddington limit only where needed.

Let us assume an external mass input rate $\dot{M}_\textup{o}$ at
the outer boundary $s_\textup{o}$.
As the critical accretions rates in the outer part of a thin
$\alpha$-disc become quite large, we expect the disc there to be
sub-critical for any reasonable
$\dot{M}_\textup{o}$. But, at a certain radius $s_\textup{crit}$ -- which
can be calculated directly from~\eqref{eqn_result_mpcrit_thin} -- the
external mass input rate may become equal to the local Eddington
limit $\dot{M}_\textup{crit}(s_\textup{crit})$. As
$\dot{M}_\textup{crit}(s)$ is monotonically decreasing (for the
influence of the inner boundary condition, see the discussion
above), the disc will remain in a critical state for all $s \leq
s_\textup{crit}$ when assuming that mass outflows guarantee the
condition $\dot{M}(s) = \dot{M}_\textup{crit}(s)$ for $s \leq
s_\textup{crit}$.

This leads to discrepancies in the disc model, as the
$\alpha$-disc equations have been derived under the condition
$\dot{M} = \textup{const}$. This requires the application of
non-conservative disc models as presented by \citet{lipunova}. As
can be seen from her results, the disc
equations differ only slightly from that of a conservative disc
model with constant accretion rate. This is also what we expect from
comparing the time scales: The disc will adjust to changes in the
accretion rate in about the hydrostatic timescale $t_\textup{hyd} =
h/c_\textup{s}$, which itself is much shorter than the viscous timescale
$t_\textup{vis} = s/v_{s}$, as long as the disc remains thin.
Therefore, we can approximate an Eddington limited thin disc with a
quasi-static model.
\begin{figure}
\psfrag{s/si}{{$\textstyle s/s_\textup{i}$}}
\psfrag{scrit}{{$\textstyle s_\textup{crit}$}}
\psfrag{dM/dM_E}[c][c]{\begin{rotate}{-90} {$\textstyle
\dot{M}/\dot{M}_\textup{E}$} \end{rotate}}
\includegraphics[clip,width=\columnwidth]{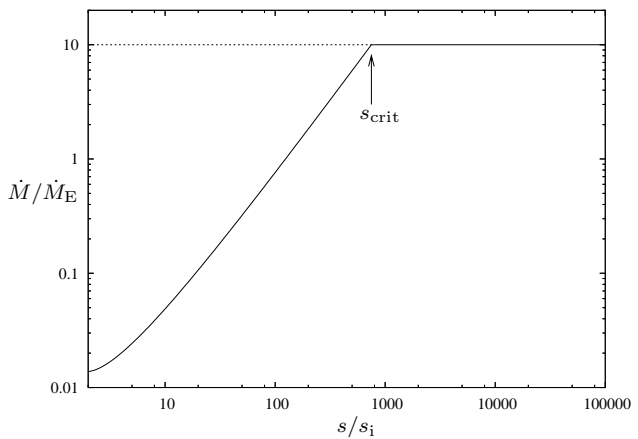}
\caption{Accretion rate for a thin disc, $\dot{M}_\textup{o} = 10 \dot{M}_\textup{E}$.
Solid/ dashed line: with/without consideration for the local (vertical)
Eddington limit}%
\label{fig_thin_disc_Mp}
\end{figure}
\begin{figure}
\psfrag{s/si}{{$\textstyle s/s_\textup{i}$}}
\psfrag{scrit}{{$\textstyle s_\textup{crit}$}}
\psfrag{h/s}[c][c]{\begin{rotate}{-90} {$\textstyle h/s$} \end{rotate}}
\includegraphics[clip,width=\columnwidth]{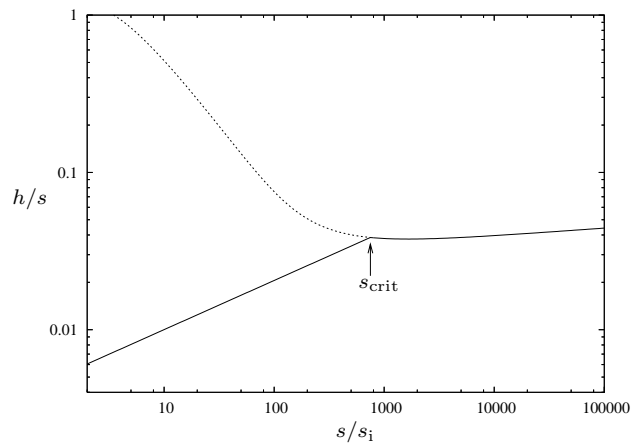}
\caption{Relative disc height $h/s$ for a thin disc with $\dot{M}_\textup{o} =
10 \dot{M}_\textup{E}$. Solid/ dashed line: with/without consideration for the
local (vertical) Eddington limit}%
\label{fig_thin_disc_h}
\end{figure}
\begin{figure}
\psfrag{s/si}{{$\textstyle s/s_\textup{i}$}}
\psfrag{scrit}{{$\textstyle s_\textup{crit}$}}
\psfrag{beta}[c][c]{\begin{rotate}{-90} {$\textstyle \beta$} \end{rotate}}
\includegraphics[clip,width=\columnwidth]{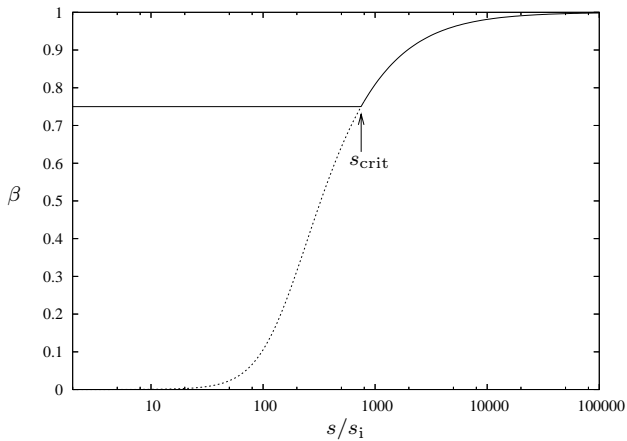}
\caption{$\beta = p_\textup{gas}/p_\textup{tot}$  for a thin disc,
$\dot{M}_\textup{o} = 10 \dot{M}_\textup{E}$. Solid/ dashed line: with/without
consideration for the local (vertical) Eddington limit}%
\label{fig_thin_disc_eta}
\end{figure}

Figures~\ref{fig_thin_disc_Mp}--\ref{fig_thin_disc_eta} show the
results for such a thin disc with $\dot{M}_\textup{o} = 10 \dot{M}_\textup{E}$.
For simplicity, we adopted simple electron scattering throughout the
whole disc and ignored the inner $2 s_\textup{i}$ because of the influence of
the boundary condition. In this case, $s_\textup{crit} \approx 750
s_\textup{i}$. As can be seen from these figures, the mass input rate to the
central object drops to about $10^{-2} \dot{M}_\textup{E}$ in the range $[2
s_\textup{i}; s_\textup{crit}]$. Also, the ratio $h/s$ follows this trend, in
contrast to the pseudo thin disc solution shown in dashed lines with
$\dot{M}(s) = \dot{M}_\textup{o} = \textup{const}$. The disc inflation in
the latter conservative case corresponds to a transition into a
radiation pressure supported disc and violates the assumptions of
geometrical thinness. On the other hand, this condition of negligible
relative height of the disc holds for an Eddington limited thin
disc, which is mainly supported by gas pressure.

The total disc luminosity, given by~\eqref{eqn_def_total_lum},
becomes $0.15 L_\textup{E}$ for the Eddington limited case and $10 L_\textup{E}$ for
the pseudo thin disc case. The total disc masses~\eqref{eqn_def_M_disc} almost equal each other,
$M_\textup{disc} \approx 10^{-6} M$ ($M$: mass of the central object),
reflecting the fact of a simple ``puffing up'' of the pseudo thin disc.
\section{Discussion}\label{sec_discussion}
In order to interpret our results correctly, we have to keep in mind
that the disc quantities presented in
Sect.~\ref{sec_results_thin_discs} and~\ref{sec_results_slim_discs}
correspond to discs at their respective \emph{local} vertical and
radial Eddington limits. The critical accretion rates in the outer
regions are enormous. However, such high mass input rates are only
formal solutions as they will never be reached in real systems.
The allowed mass accretion rates are decreasing constantly with
smaller radii, confirming our expectations that
the Eddington limit becomes important only in the inner part of the
disc.

In all our models, a strong correlation between the relative height
$h/s$ of the disc and the critical accretion rate $\dot{M}_\textup{crit}$ is
present. For a constant ratio of gas pressure to total pressure
$\beta$, the relation $h/s$ scales like $s^{0.6}$. From the
analysis of the radial limit, we know that with decreasing $\beta$
for a constant radius, the disc is blown up in vertical direction,
i.\,e. $h/s$ increases. Associated with this development, but less
effective, $\dot{M}_\textup{crit}$ increases too. We have also seen that for
sufficiently high $\dot{M}$ and $h/s$, the angular momentum in the
disc differs significantly from its Keplerian value over large
parts of the slim disc, allowing advective flows to dominate
the energy transport and to rise the critical accretion rates.

We can now draw a more realistic scenario for an accretion disc
system: Let us assume a given reasonable mass input rate
$\dot{M}_\textup{o}$ at the outer boundary of the disc. Doing so, we can
certainly assume the disc to be thin in its outer parts. For smaller
radii, the ratio $h/s$ will begin to increase as long as the
vertical Eddington limit is not violated and therefore no outflows
occur, see Fig.~\ref{fig_thin_disc_h}. Further inwards, two
competing effects now come into play: On the one hand, outflows may
occur and help the disc to stay in a subcritical state; on the other
hand, the thin disc may pass into a slim disc with deviations from
Keplerian rotation, advective energy transport and a changing ratio
$\beta$. In reality, both effects will be in concurrence. To answer
this question, extended time-dependent disc models have to be investigated.
\section{Conclusion}\label{sec_conclusion}
Our work shows that the quasi-spherical treatment of accretion
discs by applying the classical Eddington limit does not hold as
a good approximation. The allowed mass accretion rates differ
significantly from the stellar case and also depend strongly on the disc model,
while the disc luminosity only weakly exceeds the classical Eddington value.

We have seen that the inner domain of an accretion disc holds the
key position for the evolution and also for the observational
appearance of highly accreting systems: The final growth
rate of the central object, the total amount of matter that has to be expelled
from the disc and also most its luminosity are determined
by it. Numerous {radiation hydrodynamic} simulations (see \citet*{ohsuga_2005} for one example)
show that outflows in the middle part of the disc get stuck and may
therefore be driven back to the disc, while mass ejections in the
inner part lead to high velocities and gravitationally unbound
{streams} of previously accreted material.

It is broadly accepted that photon trapping effects, which are not
included in the slim disc model~\citep*{ohsuga_2002}, alter the emerging
luminosity and therefore may change our results in a way that
even higher critical accretion rates are possible. Also, the effect
of further energy transport mechanisms, like convection and
heat conduction, and non-stationarity have to be investigated
in a more sophisticated discussion.
\section*{Acknowledgments}
We would like to thank Dr Ken Ohsuga and Profs {\referee G\"{u}nther Hasinger,} Shin Mineshige and Rainer Wehrse for
helpful comments and discussions. This work was supported by the the International
Max Planck Research School for Astronomy and Cosmic Physics at the
University of Heidelberg {\referee and the Max-Planck-Institut f\"{u}r extraterrestrische Physik, Garching}.

\begin{appendix}
\section{Vertical Eddington limit in slim accretion discs}\label{app_slim_vert}
The parameterization of the pressure allows us to
reduce all thermodynamics to the parameter $\beta$:
\bea
T_\textup{c}   &=& \beta \cdot \frac{m_\textup{H} \Omega_\textup{K}^2 h^2}{2 k_\textup{B} B_1}\label{eqn_T_beta_app}\\
p_\textup{tot}&=& \frac{\beta^4}{1-\beta} \cdot \frac{\sigma}{12 c}
    \cdot \frac{m_\textup{H}^4}{k_\textup{B}^4} \cdot \frac{\Omega_\textup{K}^8 h^8}{B_1^4}\label{eqn_p_beta_app}\\
\rho &=& \frac{\beta^4}{1-\beta} \cdot \frac{\sigma}{12 c} \cdot \frac{m_\textup{H}^4}{k_\textup{B}^4} \cdot
\frac{\Omega_\textup{K}^6 h^6}{B_1^3}
\label{eqn_rho_beta_app}
\eea
We use the result~\eqref{eqn_beta_slim_vert}, $\beta = \mbox{
const.}$, together with equations (3)--(13) of \citet{abra_slim}.
More precisely, we substitute all unknowns except of $h$ and $l$ in
the radial momentum equation~\citep[(8)]{abra_slim} and the energy
equation~\citep[(12)]{abra_slim} by using the remaining disc
equations and the Eddington condition~\eqref{eqn_beta_slim_vert}.
Note that~(7) contains an error in the second term in parentheses:
The factor $(4-\beta)$ has to be replaced by $(4-3\beta)$ \citep{jaros_1}.

This leads us to two differential equations for the disc height $h$
and the angular momentum $l$, namely
\bea
\hspace*{-20pt}&&\frac{8 \Omega_\textup{K} h}{B_1} \cdot \tdif{(\Omega_\textup{K}
h)}{s} + \left(\frac{l_\textup{K}^2}{s^3} -
    \frac{l^2}{s^3}\right) =\label{eqn_slim_vert_dgl1}\\
\hspace*{-20pt}&&\qquad- \frac{\alpha \Omega_\textup{K}^3 h^3 s}{B_1 B_2 (l-l_0)^2}\cdot\left\{\Omega_\textup{K} h
+ 2s\tdif{(\Omega_\textup{K} h)}{s}
- \frac{\Omega_\textup{K} h s}{(l-l_0)} \tdif{l}{s}\right\}\nn
\eea
and
\bea
\hspace*{-20pt}&&\frac{(1-\beta)^2}{\beta^4} \cdot \frac{12 B_2
B_1^3 c^2}{\alpha \sigma \kappa} \cdot
\frac{k_\textup{B}^4}{m_\textup{H}^4} \cdot \frac{(l-l_0)}{\Omega_\textup{K}^6 h^8 s} =\label{eqn_slim_vert_dgl2}\\
\hspace*{-20pt}&&\qquad(l-l_0)\cdot \left(\frac{2l}{s^3} - \frac{1}{s^2} \tdif{l}{s}\right)
- 3\beta \frac{B_3}{B_1} \Omega_\textup{K} h \tdif{(\Omega_\textup{K} h)}{s}\,.\nn
\eea
Herein, $l_\textup{K}$ (resp. $\Omega_\textup{K}$) stands for Keplerian azimuthal
motion and $l_0$ is given by the boundary condition of a vanishing
viscous torque at the inner boundary $s_\textup{i}$ of the accretion disc,
$l_0 = l_\textup{K}(s_\textup{i})$. The critical accretion rate can be calculated from
the solutions of~\eqref{eqn_slim_vert_dgl1}
and~\eqref{eqn_slim_vert_dgl2} by
\bea
\dot{M}_\textup{crit}(s)&=&\frac{4 \pi s^2 \alpha h p_\textup{tot}}{l-l_0}\nn\\
&=&\frac{\beta^4}{1-\beta} \cdot \frac{\alpha \sigma \pi}{3 c B_1^4}
    \cdot \frac{m_\textup{H}^4}{k_\textup{B}^4} \cdot \frac{\Omega_\textup{K}^8 h^9 s^2}{l-l_0}\,.
\label{eqn_slim_vert_Mp}
\eea
%
%
\section{Radial Eddington limit in slim accretion discs}\label{app_slim_rad}
Again, we use the $\beta$-description for the disc's thermodynamics, which
yields to \eqref{eqn_T_beta_app}--\eqref{eqn_rho_beta_app}. The missing
equation to be able to determine $\dot{M}_\textup{crit}$ from the system is now given by
$v_{s} = A c_\textup{s}$ instead of the Eddington condition~\eqref{eqn_beta_slim_vert}.
From the disc equations and from the first law of thermodynamics, we already
derived~\eqref{eqn_slim_rad_GL_1}:
\[
h = B_2 \sqrt{B_1
} \cdot \frac{A}{\alpha} \cdot \left(\frac{l}{l_\textup{K}} - \frac{l_0}{l_\textup{K}}\right) s
\eqno{\eqref{eqn_slim_rad_GL_1}}
\]
As we have seen in Sect.~\ref{sec_thin_radial}, we cannot assume
$\beta = \mbox{const.}$ in this case. The radial momentum equation
and the energy equation lead to two differential equations for $h$,
$l$ and $\beta$, namely
\bea
\hspace*{-20pt}&&\tdif{}{s}\ln\left(\frac{\beta^4}{(1-\beta)}\right)
+ \left(8 + \frac{2 \alpha^2 s^2 \Omega_\textup{K}^2 h^2}{B_1 B_2^2
(l-l_0)^2}\right)
\cdot \tdif{\ln(\Omega_\textup{K} h)}{s} =\nn\\
\hspace*{-20pt}&&\frac{B_1 s}{\Omega_\textup{K}^2 h^2} (\Omega^2 - \Omega_\textup{K}^2) +
\frac{\alpha^2 s^2 \Omega_\textup{K}^2 h^2}{B_1 B_2^2 (l-l_0)^2} \cdot \frac{\left(l
+ l_0 + s^3 \tdif{\Omega}{s}\right)}{(l-l_0)\cdot s}
\label{eqn_slim_rad_dgl2}
\eea
and
\bea
\hspace*{-20pt}&&\frac{B_2 B_1^3 12 c^2}{\kappa \sigma
\Omega_\textup{K}^6 h^8 s \alpha} \cdot \frac{k_\textup{B}^4}{m_\textup{H}^4} \cdot
\frac{(1-\beta)^2}{\beta^4}
+ \tdif{\Omega}{s} =\label{eqn_slim_rad_dgl3}\\
\hspace*{-20pt}&&\quad\frac{B_3 \Omega_\textup{K}^2 h^2}{B_1 (l-l_0)} \cdot
\left\{\frac{1}{2}\left(\frac{3 \beta^2 + 3\beta - 8}{\beta(1-\beta)}\right)\tdif{\beta}{s}
- \frac{3\beta}{\Omega_\textup{K} h} \tdif{(\Omega_\textup{K} h)}{s}\right\}\,.\nn
\eea
The critical accretion rate is again given
by~\eqref{eqn_slim_vert_Mp}. Together
with~\eqref{eqn_slim_rad_GL_1}, this system can be solved
numerically.
\end{appendix}
\label{lastpage}

\begin{thebibliography}{88.}

\bibitem[\protect\citeauthoryear{Abramowicz, Calvani \&\ Nobili}{Abramowicz et al.}{1980}]{abra_thick}
Abramowicz M.A., Calvani M., Nobili L., 1980, ApJ, 242, 772

\bibitem[\protect\citeauthoryear{Abramowicz et al.}{1988}]{abra_slim} Abramowicz M.A., Czerny B.,
Lasota J.P., Szuzkiewicz E., 1988, ApJ, {332}, 646

\bibitem[\protect\citeauthoryear{Begelman}{1978}]{begelman_1978} Begelman M.C.,
1978, MNRAS, {184}, 53

\bibitem[\protect\citeauthoryear{Begelman}{2002}]{begelman_2002} Begelman M.C.,
2002, ApJ, {568}, L97

{\referee \bibitem[\protect\citeauthoryear{Bell \&\ Lin}{1994}]{bell_1994} Bell K.R., Lin D.N.C.,
1994, ApJ, {427}, 987
}

\bibitem[\protect\citeauthoryear{Collin \&\ Kawaguchi}{2004}]{collin_2004}
Collin S., Kawaguchi T., 2004, A\& A, {426}, 797

\bibitem[\protect\citeauthoryear{Duschl, Strittmatter  \&\ Biermann}{Duschl et al.}{2000}]{duschl_strittmatter_biermann}
Duschl W.J., Strittmatter P.A, Biermann P.L., 2000,
A\&A, {357}, 1123

\bibitem[\protect\citeauthoryear{Fukue}{2000}]{fukue_1} Fukue J.,
2000, PASJ, {52}, 829

\bibitem[\protect\citeauthoryear{Fukue}{2004}]{fukue_2} Fukue J.,
2004, PASJ, {56}, 569

\bibitem[\protect\citeauthoryear{Gammie}{1998}]{gammie_1998}
Gammie C.F., 1998, MNRAS, {297}, 929

\bibitem[\protect\citeauthoryear{Heinzeller}{2005}]{heinzeller} Heinzeller D.,
2005, Diploma thesis (in german), Univ. Heidelberg

\bibitem[\protect\citeauthoryear{Jaroszy{\'n}ski, Abramowicz \&\ Paczy{\'n}ski}{Jaroszy{\'n}ski et al.}{1980}]{jaros_1}
Jaroszy{\'n}ski M., Abramowicz M.A., Paczy{\'n}ski B., 1980, Acta Astronomica, {30}, 1

\bibitem[\protect\citeauthoryear{Kato, Fukue \&\ Mineshige}{Kato et al.}{1998}]{kato}
Kato S., Fukue J., Mineshige S., 1998, Black-Hole Accretion Disks. Kyoto Univ. Press, Kyoto

\bibitem[\protect\citeauthoryear{Kohri \&\ Mineshige}{1998}]{kohri} Kohri K., Mineshige S.,
2002, ApJ, {577}, 311

\bibitem[\protect\citeauthoryear{Lipunova}{1999}]{lipunova}
Lipunova G.V., 1999, Astronomy Letters, {25}, 508

\bibitem[\protect\citeauthoryear{Meyer}{2005}]{meyer_2005}
Meyer F., 2005, in Merloni A., Nayakshin S., Sunyaev R.A., eds,
Proc. of the MPA/ESO/MPE/USM Joint Astronomy Conference,
Growing black holes: accretion in a cosmological context.
Garching, Germany, p. 311

\bibitem[\protect\citeauthoryear{Ohsuga et al.}{2002}]{ohsuga_2002} Ohsuga, K.,
Mineshige S., Mori M., Umemura M., 2002, ApJ,  {574}, 315

\bibitem[\protect\citeauthoryear{Ohsuga et al.}{2005}]{ohsuga_2005} Ohsuga, K.,
Mori M., Nakamoto T., Mineshige S., 2005, ApJ, {628}, 368

\bibitem[\protect\citeauthoryear{Paczy{\'n}ski \&\ Wiita}{1980}]{paczynski} Paczy{\'n}ski B., Wiita P.J.,
1980, A\&A, {88}, 23

\bibitem[\protect\citeauthoryear{Paczy{\'n}ski \&\ Bisnovatyi-Kogan}{1981}]{paczynski_2} Paczy{\'n}ski B.,
Bisnovatyi-Kogan, G., 1981, AcA, {31}, 283

\bibitem[\protect\citeauthoryear{Ruszkowski \&\ Begelman}{2003}]{ruszkowski_2003}
Ruszkowski M., Begelman M.C., 2003, ApJ, {586}, 384

\bibitem[\protect\citeauthoryear{Shakura \&\ Sunyaev}{1973}]{shakura_sunyaev} Shakura N.I.,
Sunyaev R.A., 1973, A\& A, {24}, 337--355

\bibitem[\protect\citeauthoryear{Turner et al.}{2005}]{turner_2005}
Turner N.J., Blaes O.M., Socrates A., Begelman M.C., Davis S.W., 2005, ApJ, {624}, 267

\bibitem[\protect\citeauthoryear{Watarai \&\ Fukue}{1999}]{watarai} Watarai K.-y., Fukue J.,
1999, PASJ, {51}, 725

\bibitem[\protect\citeauthoryear{Yokosawa \&\ Uematsu}{2004}]{yokosawa} Yokosawa M., Uematsu S.,
2004, Progr. Theoret. Phys. Supp., {155}, 449
\end{thebibliography}
\end{document}